\documentclass[sigchi,screen]{acmart}

\pdfoutput=1

\makeatletter                   
\def\mdseries@tt{m}             
\makeatother                    

\usepackage[utf8]{inputenc}

\usepackage{centernot}
\usepackage{hhline}

\usepackage{siunitx}
\usepackage{textcomp}
\usepackage{caption}
\usepackage{pst-func}
\usepackage{amssymb}
\usepackage{graphicx}
\usepackage{paralist}
\usepackage{environ}
\usepackage{pgfplots}
\usepackage{tikz}
\usetikzlibrary{shadows}
\usepackage{mathtools}
\usepackage{array,multirow}
\usepackage{algorithm}
\usepackage{threeparttable}

\usepackage{tabu}

\usepackage{pgfplotstable}
\usepackage{filecontents}

\usepackage{arydshln}
\usepackage{etoolbox}

\usepackage{xspace} 
 
\usepackage{dirtytalk} 
\usepackage{stmaryrd} 

\usepackage[inline,shortlabels]{enumitem} 
\newlist{inlinelist}{enumerate*}{1}
\setlist*[inlinelist,1]{%
  label=(\roman*),
}

\usepackage{breakurl}           

\usepackage{xifthen}  
\usepackage{ifthen}

\usepackage{nicefrac}

\usepackage{hyperref}
\usepackage{cleveref}
\usepackage{bigints}

\usetikzlibrary{pgfplots.dateplot}
\usetikzlibrary{matrix,chains,positioning,decorations.pathreplacing}
\usetikzlibrary{patterns,calc,fit,arrows}
\usetikzlibrary{shapes.geometric}

\usepackage[final,nomargin,inline,index]{fixme} 
\fxusetheme{color}

\FXRegisterAuthor{bart}{anbart}{\color{magenta} {\underline{bart}}}
\FXRegisterAuthor{tizi}{antizi}{\color{red} {\underline{tizi}}}
\FXRegisterAuthor{livio}{anlivio}{\color{blue} {\underline{livio}}}
\FXRegisterAuthor{sergio}{ansergio}{\color{green} {\underline{sergio}}}

\hypersetup{final} 

\usepackage{listings}
\definecolor{listingBG}{HTML}{FFFFCB}%
\definecolor{listingFrame}{HTML}{BBBB98}%
\definecolor{listingLineno}{rgb}{0.5,0.5,1.0}%

\definecolor{LightGrey}{rgb}{0.975,0.975,0.975}

\lstdefinelanguage{btm}{
	commentstyle=\color{Gray},
	morecomment=[l]{//},
	morecomment=[s]{/*}{*/},
	classoffset=0,
        escapechar=\$,
	morekeywords={transaction,input,output,key,network,package},
	keywordstyle=\color{Blue}\bfseries,
	classoffset=1,
	morekeywords={sig,versig,fun,unit,int,string,bool,address,uint},
	keywordstyle=\color{TealBlue},
	classoffset=2,
	morekeywords={BTC,true},
	keywordstyle=\color{Plum}\bfseries,
}

\lstdefinelanguage{solidity}{
	commentstyle=\color{Gray},
	morecomment=[l]{//},
	morecomment=[s]{/*}{*/},
	classoffset=0,
        escapechar=\$,
	morekeywords={struct,mapping,function,this,public,private,static,final,class,extends,switch,case,break,finally,try,catch,return,if,else,new},
	keywordstyle=\color{Blue}\bfseries,
	classoffset=1,
	morekeywords={unit,int,string,bool,address,uint},
	keywordstyle=\color{TealBlue},
	classoffset=2,
	morekeywords={ether,wei,finney,contract,send,throw,msg,sender,value},
	keywordstyle=\color{Plum}\bfseries,
}

\lstdefinelanguage{java}{
	escapechar=\$,
        commentstyle=\color{Gray},
	morecomment=[l]{//},
	morecomment=[s]{/*}{*/},
	morestring=[b]",
        classoffset=0,
	morekeywords={public,private,static,final,class,extends,switch,case,break,finally,try,catch,void,int,boolean,throws,throw,return,if,else,new},
	keywordstyle=\color{keyword}\bfseries
}

\newcommand{\ifempty}[3]{%
  \ifthenelse{\isempty{#1}}{#2}{#3}%
}

\newcommand{\ifdots}[3]{%
  \ifthenelse{\equal{#1}{...}}{#2}{#3}%
}

\newcommand{\hidden}[1]{}

\makeatletter
\pgfplotsset{
    /pgfplots/flexible xticklabels from table/.code n args={3}{%
        \pgfplotstableread[#3]{#1}\coordinate@table
        \pgfplotstablegetcolumn{#2}\of{\coordinate@table}\to\pgfplots@xticklabels
        \let\pgfplots@xticklabel=\pgfplots@user@ticklabel@list@x
    }
}
\makeatother

\newcommand{\USD}{\mbox{\textit{USD}}\xspace}

\newcommand{\Real}[1]{\mathrm{Real}}

\newcommand{\eg}{e.g.\@\xspace}
\newcommand{\ie}{i.e.\@\xspace}

\newcommand{\BTC}{\textup{%
  \leavevmode
  \vtop{\offinterlineskip 
    \setbox0=\hbox{B}%
    \setbox2=\hbox to\wd0{\hfil\hskip-.03em
    \vrule height .3ex width .15ex\hskip .08em
    \vrule height .3ex width .15ex\hfil}
    \vbox{\copy2\box0}\box2}}\xspace}

\crefname{appendix}{appendix}{appendices}
\Crefname{appendix}{Appendix}{Appendices}
\crefname{notation}{notation}{notations}
\Crefname{notation}{Notation}{Notations}

\definecolor{LightGrey}{rgb}{0.95,0.95,0.95}
\definecolor{keyword}{HTML}{7F0055}

\definecolor{AgricultureColor}{RGB}{203,172,136}
\definecolor{DemocracyColor}{RGB}{255,165,0}
\definecolor{IdentityColor}{RGB}{142,158,246}
\definecolor{EducationColor}{RGB}{255,151,255}
\definecolor{EnvironmentColor}{RGB}{88,162,121}
\definecolor{FinancialColor}{RGB}{173,216,230}
\definecolor{HealthColor}{RGB}{255,92,174}
\definecolor{HumanColor}{RGB}{237,182,163}
\definecolor{LandColor}{RGB}{248,233,233}
\definecolor{PhilanthropyColor}{RGB}{168,215,182}

\makeatletter

\tikzstyle{chart}=[
legend label/.style={font={\scriptsize},anchor=west,align=left},
legend box/.style={rectangle, draw, minimum size=5pt},
axis/.style={black,semithick,->},
axis label/.style={anchor=east,font={\tiny}},
]

\tikzstyle{bar chart}=[
chart,
bar width/.code={
	\pgfmathparse{##1/2}
	\global\let\bar@w\pgfmathresult
},
bar/.style={very thick, draw=white},
bar label/.style={font={\bf\small},anchor=north},
bar value/.style={font={\footnotesize}},
bar width=.75,
]

\tikzstyle{pie chart}=[
chart,
slice/.style={line cap=round, line join=round, very thick,draw=white},
pie title/.style={font={\bf}},
slice type/.style 2 args={
	##1/.style={fill=##2},
	values of ##1/.style={}
}
]

\pgfdeclarelayer{background}
\pgfdeclarelayer{foreground}
\pgfsetlayers{background,main,foreground}

\newcommand{\pie}[3][]{
	\begin{scope}[#1]
		\pgfmathsetmacro{\curA}{90}
		\pgfmathsetmacro{\r}{1}
		\def\c{(0,0)}
		\node[pie title] at (90:1.3) {#2};
		\foreach \v/\s in{#3}{
			\pgfmathsetmacro{\deltaA}{\v/100*360}
			\pgfmathsetmacro{\nextA}{\curA + \deltaA}
			\pgfmathsetmacro{\midA}{(\curA+\nextA)/2}
			
			\path[slice,\s] \c
			-- +(\curA:\r)
			arc (\curA:\nextA:\r)
			-- cycle;
			\pgfmathsetmacro{\d}{max((\deltaA * -(.5/50) + 1) , .5)}
			
			\begin{pgfonlayer}{foreground}
				\path \c -- node[pos=\d,pie values,values of \s]{$\v\%$} +(\midA:\r);
			\end{pgfonlayer}
			
			\global\let\curA\nextA
		}
	\end{scope}
}

\newcommand{\legend}[2][]{
	\begin{scope}[#1]
		\path
		\foreach \n/\s in {#2}
		{
			++(1.5,0cm) node[\s,legend box] {} +(0.1cm,0) node[legend label] {\n}
		}
		;
	\end{scope}
}

\newcommand{\urldataset}{https://goo.gl/Erfm86}

\newcommand{\urlgithub}{https://github.com/blockchain-unica/social-good}

\newcommand{\totProjects}{\mbox{120}\xspace}

\newcommand{\percentProjectsPhilanthropy}{\mbox{31}\%\xspace}

\newcommand{\receiveAssetToken}{16}
\newcommand{\receiveAssetFiat}{17}
\newcommand{\receiveAssetCrypto}{31}

\newcommand{\receiveAssetNo}{11}

\newcommand{\receiveReasonDonation}{36}
\newcommand{\receiveReasonProfit}{14}

\newcommand{\moneyRecipientPerson}{13}
\newcommand{\moneyRecipientCharity}{8}
\newcommand{\moneyRecipientProject}{26}

\newcommand{\recipientChoiceSender}{41}
\newcommand{\recipientChoicePlatform}{11}
\newcommand{\recipientChoiceVoting}{4}

\newcommand{\impactFeedbackYes}{15}
\newcommand{\impactFeedbackNo}{34}

\newcommand{\moneyTransferAutomatic}{18}
\newcommand{\moneyTransferManual}{43}

\newcommand{\totProjectsFunding}{\mbox{35}\xspace}

\newcommand{\totProjectsFundingOther}{\mbox{8}\xspace}

\newcommand{\usdProjectsFundingOther}{\mbox{\USD 65.3} millions\xspace}

\newcommand{\totProjectsFundingICOclosed}{\mbox{34}\xspace}

\newcommand{\totProjectsFundingICOnonzero}{\mbox{26}\xspace}

\newcommand{\totProjectsFundingICOany}{\mbox{45}\xspace}

\newcommand{\usdProjectsFundingICO}{\mbox{\USD 346.4} millions\xspace}

\newcommand{\totProjectsFundingICOboth}{\mbox{17}\xspace}

\newcommand{\totProjectsFundingICOinsuccess}{\mbox{6}\xspace}

\newcommand{\totProjectsFundingICOsuccess}{\mbox{11}\xspace}

\newcommand{\totProjectsGithub}{\mbox{38}\xspace}

\newcommand{\totProjectsGithubAlive}{\mbox{23}\xspace}

\newcommand{\percentProjectsGithubAlive}{\mbox{60\%}\xspace}

\newcommand{\totProjectsWhitepaper}{\mbox{62}\xspace}

\newcommand{\totProjectsGithubWhitepaper}{\mbox{28}\xspace}

\newcommand{\totProjectsFundedGithubWhitepaper}{\mbox{15}\xspace}

\newcommand{\totProjectsInfoBlockchain}{\mbox{100}\xspace}

\newcommand{\totProjectsMoreBlockchain}{\mbox{12}\xspace}

\newcommand{\totProjectsBlockchainEthereum}{\mbox{67}\xspace}

\newcommand{\totProjectsBlockchainCustom}{\mbox{18}\xspace}

\newcommand{\totProjectsActualDiffersPredicted}{\mbox{8}\xspace}

\newcommand{\totProjectsArchitecturePermissionlessPermissioned}{\mbox{5}\xspace}

\makeatletter
\renewcommand\paragraph{\@startsection{paragraph}{4}{\z@}%
  {2.25ex \@plus 1ex \@minus .2ex}%
  {-0.75em}%
  {\normalfont\normalsize\bfseries}}
\makeatother

\copyrightyear{2018}
\acmYear{2018}
\setcopyright{acmlicensed}
\acmConference[Goodtechs '18]{International Conference on Smart Objects and Technologies for Social Good}{November 28--30, 2018}{Bologna, Italy}
\acmBooktitle{International Conference on Smart Objects and Technologies for Social Good (Goodtechs '18), November 28--30, 2018, Bologna, Italy}
\acmPrice{15.00}
\acmDOI{10.1145/3284869.3284881}
\acmISBN{978-1-4503-6581-9/18/11}

\renewcommand\footnotetextcopyrightpermission[1]{} 
\begin{document}
\sloppy                         

\newcommand{\mytitle}{Blockchain for social good: a quantitative analysis}
\title{\mytitle}

\author{Massimo Bartoletti}
\affiliation{\institution{University of Cagliari}}
\email{bart@unica.it}

\author{Tiziana Cimoli}
\affiliation{\institution{University of Cagliari}}
\email{t.cimoli@unica.it}

\author{Livio Pompianu}
\affiliation{\institution{University of Cagliari}}
\email{livio.pompianu@unica.it}

\author{Sergio Serusi}
\affiliation{\institution{University of Cagliari}}
\email{sergio.serusi@unica.it}

\begin{abstract}
  The rise of blockchain technologies has given a boost 
  to social good projects,
  which are trying to exploit 
  various characteristic features of blockchains:
  the quick and inexpensive transfer of cryptocurrency,
  the transparency of transactions,
  the ability to tokenize any kind of assets,
  and the increase in trustworthiness due to decentralization.
  However, the swift pace of innovation in blockchain technologies,
  and the hype that has surrounded their ``disruptive potential'',
  make it difficult to understand whether these technologies are
  applied correctly, and what one should expect when trying to
  apply them to social good projects.
  This paper addresses these issues, by systematically analysing a 
  collection of \totProjects blockchain-enabled social good projects.
  Focussing on measurable and objective aspects,
  we try to answer various relevant questions:
  which features of blockchains are most commonly used?
  Do projects have success in fund raising?
  Are they making appropriate choices on the blockchain architecture?
  How many projects are released to the public,
  and how many are eventually abandoned?
\end{abstract}



\maketitle

\thispagestyle{empty}
\pagestyle{empty}

\section{Introduction}

In the last few years there has been a steady increase
of interest in blockchain technologies.
This is witnessed --- among the other things --- 
by the venture capital funding of billions dollars 
in blockchain start-ups~\cite{Friedlmaier18hicss},
the proliferation of open-source projects~\cite{Deloitte17github},
and the interest of major ICT and consultancy 
companies~\cite{Deloitte16democratized} and 
national governments~\cite{UK16report}. 

Among the various fields where blockchain technologies are believed to
have an impact, 
social good is among those that are generating the 
greatest expectations~\cite{Galen18rippleworks,Pisa18hypevsreality}.
However, it is not easy to foresee whether these expectations will be met.
On the one side, the evangelists of blockchain technologies think that
\emph{``blockchain will touch, if not disrupt, every
major industry and will even alter the way that people and societies
interact''}~\cite{Galen18rippleworks}.
On the other side, blockchain skeptics believe that 
the flaunted ``disruptive potential'' of these technologies in only hype,
as no convincing use case has been found yet~\cite{Stinchcombe18tenyears}.
From a strictly technical perspective, these skeptics find support in that
every blockchain use case can also be implemented \emph{without} a blockchain:
indeed, the added value of blockchains is that 
they can weaken the trust assumptions of an application.
For instance, Bitcoin --- the first blockchain-enabled cryptocurrency ---
implements a globally-agreed ledger of currency transactions, 
which is maintained by a P2P network~\cite{Bonneau15ieeesp}. 
Unlike the previous generation of cryptocurrencies, 
which required a trusted authority to maintain the ledger,
the only assumption underlying the security of Bitcoin is that nodes have a rational behaviour, 
\ie their choices are driven by economic incentives. 
This decentralization of trust 
--- from a single authority to a network of mutually distrusted nodes ---
is the real potential of blockchains (at least, of their ``permissionless'' incarnations).

Although decentralization could play a role in determining the success of social good applications,
there are still no objective data on how the projects 
that have been proposed over the last years are actually behaving.
The most recent reports on blockchain for social good
are quite limited regarding objective data:
they just describe some use cases, 
trying to motivate the applicability of blockchains~\cite{Pisa18hypevsreality},
or provide statistics about social good projects
on the basis of interviews to their proposers~\cite{Galen18rippleworks}
\footnote{These interviews are mainly focussed on \emph{subjective} data, asking 
\eg, ``how does your initiative use blockchain, and why is blockchain a good technology for this problem?'',
or ``in what time frame do you think you will see meaningful impact from your blockchain initiative?''.}.

Objective measures on social good projects could help 
to separate the hype from the reality.
The web already makes available several sources of measurable data:
for instance, the projects websites, their code repositories,
the crowdfunding and ICO rating platforms,
besides blogs and social networks. 







  \paragraph{Contributions}

This paper is a quantitative analysis of blockchain-enabled social good projects,
based on publicly available data. 
In summary, our main contributions are:
\begin{enumerate}


\item a public dataset of \totProjects blockchain-enabled social good projects
  (\href{\urldataset}{\urldataset}),
  containing all the data needed to reproduce the analyses developed 
  in our survey;

\item an open-source repository of projects descriptions
  (\href{\urlgithub}{\urlgithub});

\item a study of the distribution of social good projects among 
  different impact sectors;

\item an analysis of the main features of projects,
  focussed at discovering how they exploit blockchain technologies;

\item an estimate of how, and how much, projects have gathered investments,
  and of the success of fund raising campaigns 
  with respect to the expectations;

\item a systematic study of the architectural choices made by projects,
  and a comparison between the actual type of blockchain chosen and 
  the one predicted by the decision making process in~\cite{Wust18cvcbt};

\item an evaluation of the status of projects, 
  taking into account both their online channels and 
  the activity on their code repositories,
  aimed at measuring the successful deployment and the mortality of projects.

\end{enumerate}

\section{Collection of social good projects}
\label{sec:collection}

We have crawled the web for articles on social good projects,
focussing in particular on websites which rate ICOs
(like \eg \href{https://icobench.com/icos}{icobench.com}).
By manually filtering these results, 
we have collected \totProjects social good projects based on blockchain. 
Our criteria for deciding whether or not a project 
should be in the collection are the following:
\begin{enumerate}
\item \label{item:collection:blockchain}
  the project declares to use a blockchain, of any kind;
\item \label{item:collection:UN}
  the goals of the project must be coherent with 
  the \href{https://sustainabledevelopment.un.org/}{UN Sustainable Development Goals};
\item \label{item:collection:socialgood}
  social good should be the preeminent goal of the project.
\end{enumerate}

We have chosen to keep in our collection also the projects
that have already been abandoned, 
since we want to measure the mortality of projects.
We have also kept in our collection the projects 
that are already operational without any blockchain, 
but that have planned to switch to blockchain in the near future
(this is the case \eg of \href{http://govtechfund.com/2017/11/neighborly-bringing-blockchain-to-municipal-bonds/}{Neighborly}).
One of the most delicate choices we had to make 
concerns the projects in the health sector and in the energy sector.
Although health and energy are coherent with the 
UN Sustainable Development Goals,
many project in these sectors are preeminently business-oriented, 
thus violating our third criterion.

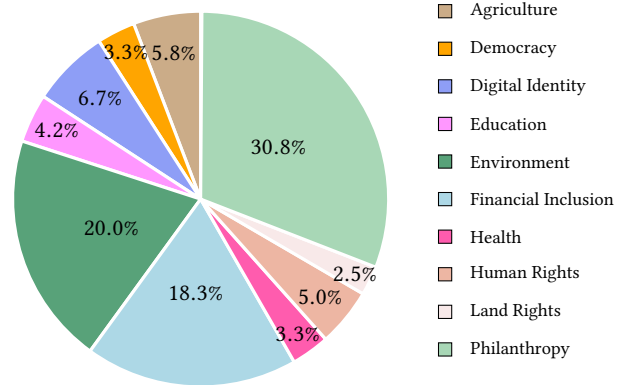
\begin{figure}
  \centering
  \scriptsize
  \scalebox{1}{
    \begin{tikzpicture}
      [
      pie chart,
      slice type={Agriculture}{AgricultureColor},
      slice type={Democracy}{DemocracyColor},
      slice type={Digital Identity}{IdentityColor},
      slice type={Education}{EducationColor},
      slice type={Environment}{EnvironmentColor},
      slice type={Financial Inclusion}{FinancialColor},
      slice type={Health}{HealthColor},
      slice type={Human Rights}{HumanColor},
      slice type={Land Rights}{LandColor},
      slice type={Philanthropy}{PhilanthropyColor},
      pie values/.style={font={\small}},
      scale=2.5
      ]
      
      \pie{}{
        5.8/Agriculture, 
        3.3/Democracy, 
        6.7/Digital Identity, 
        4.2/Education, 
        20.0/Environment, 
        18.3/Financial Inclusion, 
        3.3/Health, 
        5.0/Human Rights, 
        2.5/Land Rights, 
        30.8/Philanthropy}
      
      \legend[shift={(-0.2cm, 1.0cm)}]{{Agriculture}/Agriculture}
      \legend[shift={(-0.2cm, 0.8cm)}]{{Democracy}/Democracy}
      \legend[shift={(-0.2cm, 0.6cm)}]{{Digital Identity}/Digital Identity} 
      \legend[shift={(-0.2cm, 0.4cm)}]{{Education}/Education}
      \legend[shift={(-0.2cm, 0.2cm)}]{{Environment}/Environment}
      \legend[shift={(-0.2cm, 0.0cm)}]{{Financial Inclusion}/Financial Inclusion}
      \legend[shift={(-0.2cm, -0.2cm)}]{{Health}/Health}
      \legend[shift={(-0.2cm, -0.4cm)}]{{Human Rights}/Human Rights}
      \legend[shift={(-0.2cm, -0.6cm)}]{{Land Rights}/Land Rights}
      \legend[shift={(-0.2cm, -0.8cm)}]{{Philanthropy}/Philanthropy}
    \end{tikzpicture}
  }
  \caption{Projects by social impact sector.}
  \label{fig:impact-sectors}
\end{figure}

After the collection phase,
we have analysed the projects to perform a first categorization, 
based on their social impact sector.
To this purpose we have followed 
the taxonomy in~\cite{Galen18rippleworks},
so to be able to compare their results with those in our survey.
Our categorization is shown in~\Cref{fig:impact-sectors},
from where we see that the most populated category is ``Philanthropy'' (\percentProjectsPhilanthropy)%
\footnote{In contrast with~\cite{Galen18rippleworks}, where ``Health'' is in first place.
This discrepancy is due to our choice to exclude business-oriented health projects.},
which includes \eg charity donation platforms.
The ``Environment'' category includes projects whose main purpose is to improve the quality of the environment,
as well as projects aimed at optimizing the usage and distribution of energy.
``Financial Inclusion'' comprises \eg software platforms for remittances and for micro-loans.

\section{What do  social good projects do?}
\label{sec:taxonomy}

We now isolate some archetypal features of social good projects, 
and we measure them for those in our collection.
Since most projects allow for exchanging some kind of asset, 
we focus on features which impact the financial side%
\footnote{In our measures we omit the projects for which it was not possible
to infer the values of the features from the online documentation.}. %

We start by studying which kind of asset the projects accept as input, if any. 
The results of our analysis are displayed by the following diagram.
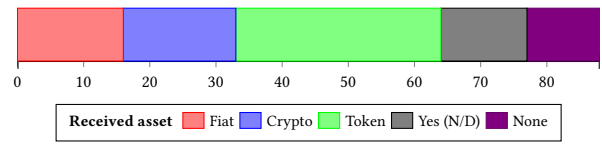
\begin{center}
  \begin{tikzpicture}
    \begin{axis}[xbar stacked,
      legend style={
        at={(0.5,-0.15)}, 
        anchor=north,
        legend columns=-1,
        font=\tiny,
      },
      xmin=0,
      height = 3cm,
      width = 1.1*\columnwidth,
      bar width=20pt,
      hide y axis,
      axis x line=bottom,
      axis line shift=-10pt,
      every tick label/.append style={font=\tiny},
      axis line style={|-|}
      ]
      \addlegendimage{empty legend};
      \pgfplotstableread[col sep=comma]{feature-in-medium.csv}\mytable
      \addplot[red,fill=red!50] table [x index=0] {\mytable};
      \addplot[blue,fill=blue!50] table [x index=1] {\mytable};
      \addplot[green,fill=green!50] table [x index=2] {\mytable};
      \addplot table [x index=3] {\mytable};
      \addplot table [x index=4] {\mytable};
      \addlegendentry{\textbf{Received asset\;}}
      \addlegendentry{Fiat}
      \addlegendentry{Crypto}
      \addlegendentry{Token}
      \addlegendentry{Yes (N/D)}
      \addlegendentry{None}
    \end{axis}
  \end{tikzpicture}       
\end{center}

\smallskip
We see that of all the projects, only \receiveAssetNo\xspace 
seem not to receive any kind of asset. 
Example of these are:
\href{https://www.fluxtoken.io/}{Flux}, which gathers
environmental data to help farmers improve yields;
\href{https://poseidon.eco/}{Poseidon}, which help to
track one's carbon footprint; and
\href{handshake.tech/index.html}{Handshake}, 
which stores on the blockchain the contracts
of migrant workers, to help reduce abuse.
For the other projects, the most common ways to receive assets
are cryptocurrency (chosen by \receiveAssetCrypto\xspace projects) and 
fiat currency (\receiveAssetFiat\xspace projects).
Some of the projects manage both kinds of currency:
for instance, wallet applications
(\href{https://getchange.com/}{ChangeBank},
\href{https://bitpay.com/}{BitPay},
\href{https://www.unocoin.com/}{Unocoin},
\href{https://kora.network/}{Kora}) %
which help users to trade fiat for cryptocurrencies, and send (any)
assets to someone else, anywhere, in a short time.  
Projects of this kind target people living in unbanked regions and allow them to
receive remittances from their family without the need of a bank account.
Besides fiat and crypto, the other possibility is that the  project 
implements its own token, and uses it as a  medium for payments. 
This option has been chosen by \receiveAssetToken\xspace projects 
(\href{https://getstarted.with.pink/}{Pink},
\href{https://moedaseeds.com/}{Moeda},
\href{http://monedapar.com/}{MonedaPar},
\href{https://www.almbank.io/}{AlmBank}).
Some projects 
(\eg, \href{https://www.aidcoin.co/?lang=en}{AidCoin} and
\href{https://sureremit.co/}{SureRemit})
receive fiat or crypto, and then convert them into tokens; 
alternatively, tokens have to be purchased through exchanges 
like \href{https://www.coinbase.com/}{coinbase.com}. 
Buying tokens is not the only way to obtain them:
in some cases, tokens can be earned as reward for some behaviour,
like \eg using the application (\href{https://www.getwala.com/}{Wala}), 
attending school (\href{https://give.si/}{GiveFoundation}); 
tokens may also serve as a basic subsidy to eradicate poverty 
(\href{https://www.mannabase.com/}{Mannabase}).
 
The following diagram measures the reason why money is sent to projects.
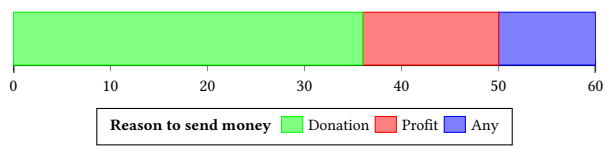
\begin{center}
  \begin{tikzpicture}
    \begin{axis}[xbar stacked,
      legend style={
        at={(0.5,-0.15)}, 
        anchor=north,
        legend columns=-1,
        font=\tiny,
      },
      xmin=0,
      height = 3cm,
      width = 1.1*\columnwidth,
      bar width=20pt,
      hide y axis,
      axis x line=bottom,
      axis line shift=-10pt,
      every tick label/.append style={font=\tiny},
      axis line style={|-|}
      ]
      \addlegendimage{empty legend};
      \pgfplotstableread[col sep=comma]{feature-reason.csv}\mytable
      \addplot[green,fill=green!50] table [x index=0] {\mytable};
      \addplot[red,fill=red!50] table [x index=1] {\mytable};
      \addplot[blue,fill=blue!50] table [x index=2] {\mytable};
      \addlegendentry{\textbf{Reason to send money\;}}
      \addlegendentry{Donation}
      \addlegendentry{Profit}
      \addlegendentry{Any}
    \end{axis}
  \end{tikzpicture}       
\end{center}

\smallskip
In the vast majority of cases (\receiveReasonDonation), 
the money is sent as a donation 
(\eg, in \href{https://bithope.org/}{BitHope},
\href{https://www.aidcoin.co/?lang=en}{AidCoin},
\href{hhttp://charity-dao.org/}{CharityDAO}).  
Indeed, projects that collect money for charities are soaring, 
since blockchain technologies allow for fast transfers of money across countries, 
with the added benefit of making the flow of money observable by donors.
Beside donations, \receiveReasonProfit\xspace projects try to couple social good with 
the possibility to make profit%
\footnote{We are not considering the mere ownership of crypto as an investment.}.
For instance,
\href{https://moedaseeds.com/}{Moeda} and
\href{https://www.ethichub.com/}{EthicHub} are platforms to crowdfund
agricultural projects, which, in case of success, will pay back investors;
\href{https://www.wetrust.io/}{WeTrust} and
\href{https://www.surco.in/}{Suretly} implement micro loans;
\href{https://bananacoin.io/}{Banancoin} and 
\href{https://cacaoshares.com/}{CacaoShares} sell tokens linked to the
harvesting of goods, with the idea that when the product is ready and
sold, the token are paid back; %
\href{http://batan.io/}{Batan} tracks carbon footprint, and assists in buying carbon credits;
\href{https://www.recycletocoin.com/}{RecicleToCoin} aims
at paying people in order to collect plastic waste, targeting the
double goal of helping people out of poverty and cleaning the environment.

The following diagram measures who is the ultimate recipient of the money sent
to social good applications:
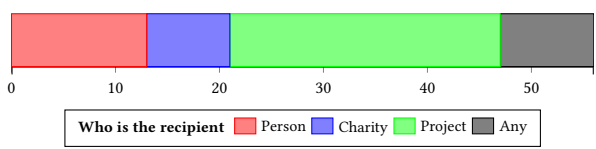
\begin{center}
  \begin{tikzpicture}
    \begin{axis}[xbar stacked,
      legend style={
        at={(0.5,-0.15)}, 
        anchor=north,
        legend columns=-1,
        font=\tiny,
      },
      xmin=0,
      height = 3cm,
      width = 1.1*\columnwidth,
      bar width=20pt,
      hide y axis,
      axis x line=bottom,
      axis line shift=-10pt,
      every tick label/.append style={font=\tiny},
      axis line style={|-|}
      ]
      \addlegendimage{empty legend};
      \pgfplotstableread[col sep=comma]{feature-recipient.csv}\mytable
      \addplot[red,fill=red!50] table [x index=0] {\mytable};
      \addplot[blue,fill=blue!50] table [x index=1] {\mytable};
      \addplot[green,fill=green!50] table [x index=2] {\mytable};
      \addplot table [x index=3] {\mytable};
      \addlegendentry{\textbf{Who is the recipient\;}}
      \addlegendentry{Person}
      \addlegendentry{Charity}
      \addlegendentry{Project}
      \addlegendentry{Any}
    \end{axis}
  \end{tikzpicture}
\end{center}

\smallskip
In \moneyRecipientPerson\xspace cases the recipient is an actual person. 
Most of these cases are  wallet applications that do remittances, and  charity
projects which aim at helping individuals. 
For instance,
\href{https://www.fastcompany.com/40500978/this-new-blockchain-project-gives-homeless-new-yorkers-a-digital-identity}{Fummy}
and \href{https://www.hypergive.com/}{Hypergive} transfer money
directly to homeless people to allow them buy food;
\href{https://give.si/}{Give} project sends money to children who attend school.
In \moneyRecipientCharity\xspace philanthropic projects
(\eg, \href{https://goodcoin.to/}{GoodCoin},
\href{https://www.humanityroad.org/bitcoin}{HumanityRoad},
\href{https://urbanarray.org/}{UrbanArray}),
funds are sent to a charity, with no further specification on how they will be spent.
In the vast majority of cases (\moneyRecipientProject), 
money is gathered to fund a specific project (that can be either for charity or for investment).
For instance, \href{https://neighborly.com/}{Neighborly} allows users to
choose the projects to invest in, among those who benefit the local community;
\href{https://www.givetrack.org/}{GiveTrack} offers a choice of different projects to which one can donate money. 

The following diagram measures who decides the recipient of money:
this can be the sender himself, the application, or a voting procedure.
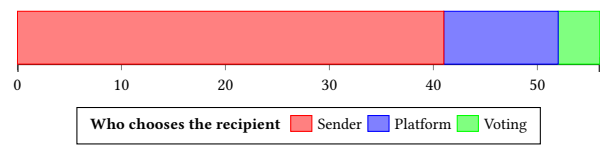
\begin{center}
  \begin{tikzpicture}
    \begin{axis}[xbar stacked,
      legend style={
        at={(0.5,-0.15)}, 
        anchor=north,
        legend columns=-1,
        font=\tiny,
      },
      xmin=0,
      height = 3cm,
      width = 1.1*\columnwidth,
      bar width=20pt,
      hide y axis,
      axis x line=bottom,
      axis line shift=-10pt,
      every tick label/.append style={font=\tiny},
      axis line style={|-|}
      ]
      \addlegendimage{empty legend};
      \pgfplotstableread[col sep=comma]{feature-chooser.csv}\mytable
      \addplot[red,fill=red!50] table [x index=0] {\mytable};
      \addplot[blue,fill=blue!50] table [x index=1] {\mytable};
      \addplot[green,fill=green!50] table [x index=2] {\mytable};
      \addlegendentry{\textbf{Who chooses the recipient\;}}
      \addlegendentry{Sender}
      \addlegendentry{Platform}
      \addlegendentry{Voting}
    \end{axis}
  \end{tikzpicture}               
\end{center}

\smallskip
Most projects (\recipientChoiceSender\xspace cases) allow senders to choose who receives their donations.
In \recipientChoicePlatform\xspace cases 
(\eg, \href{https://goodcoin.to/}{GoodCoin},
\href{https://charitytokenonline.com/}{CharityToken},
\href{http://rmblockchain.org/}{DistributeGiving})
the choice is made by the platform itself,
often without detailing their policy. %
In \recipientChoiceVoting\xspace projects
(\eg, \href{http://www.benefactory.cc/}{Benefactory},
\href{http://www.disberse.com/}{Positive Women}.
\href{http://charity-dao.org/}{CharityDAO}) 
the choice is taken after a vote among the senders.

The following diagram analyses whether who sends money to a project
can obtain some feedback on what has been accomplished with their money.
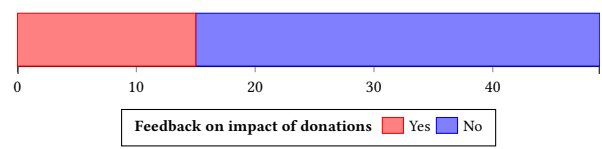
\begin{center}
  \begin{tikzpicture}
    \begin{axis}[xbar stacked,
      legend style={
        at={(0.5,-0.15)}, 
        anchor=north,
        legend columns=-1,
        font=\tiny,
      },
      xmin=0,
      height = 3cm,
      width = 1.1*\columnwidth,
      bar width=20pt,
      hide y axis,
      axis x line=bottom,
      axis line shift=-10pt,
      every tick label/.append style={font=\tiny},
      axis line style={|-|}
      ]
      \addlegendimage{empty legend};
      \pgfplotstableread[col sep=comma,header=true]{feature-feedback.csv}\mytable;
      \addplot[red,fill=red!50] table [x index=0] {\mytable};
      \addplot[blue,fill=blue!50] table [x index=1] {\mytable};
      \addlegendentry{\textbf{Feedback on impact of donations\;}}
      \addlegendentry{Yes}
      \addlegendentry{No}
    \end{axis}
  \end{tikzpicture}        
\end{center}

We observe that in the majority of projects (\impactFeedbackNo\xspace cases), 
no impact data is provided,
while \impactFeedbackYes\xspace projects provide donors with some sort of feedback.
For instance, \href{https://www.givedirectly.org/}{GiveDirectly} features 
a web platform through which the recipients of donations 
can promote their cause or provide feedback.
\href{https://sureremit.co/}{SureRemit} implements a remittance service
which allows senders to know how their money has been spent.  %
\href{ https://www.ethichub.com/}{EthicHub} implements a reputation mechanism 
which rewards farmers that repaid their loans.
\href{http://www.amply.tech/}{Amply} and
\href{https://www.educategirls.ngo/}{EducateGirls} 
are projects developed by the \href{https://ixo.foundation/}{IXO Foundation} 
to provide education for children in Africa and India;
both projects store impact data to assess the project progression. %
In the \href{https://alice.si/}{Alice} platform, 
funded projects are monitored and assessed. %
In the \href{https://www.giftcoin.org/}{GiftCoin} platform, 
projects are funded incrementally, according to assessment results.

Finally, we study whether projects implement some 
mechanism to automatically trigger payments.
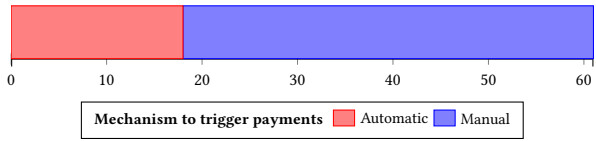
\begin{center}
  \begin{tikzpicture}
    \begin{axis}[xbar stacked,
      legend style={
        at={(0.5,-0.15)}, 
        anchor=north,
        legend columns=-1,
        font=\tiny,
      },
      xmin=0,
      height = 3cm,
      width = 1.1*\columnwidth,
      bar width=20pt,
      hide y axis,
      axis x line=bottom,
      axis line shift=-10pt,
      every tick label/.append style={font=\tiny},
      axis line style={|-|}
      ]
      \addlegendimage{empty legend};
      \pgfplotstableread[col sep=comma]{feature-transfer.csv}\mytable
      \addplot[red,fill=red!50] table [x index=0] {\mytable};
      \addplot[blue,fill=blue!50] table [x index=1] {\mytable};
      \addlegendentry{\textbf{Mechanism to trigger payments\;}}
      \addlegendentry{Automatic}
      \addlegendentry{Manual}
    \end{axis}
  \end{tikzpicture}        
\end{center}

\smallskip
The majority of projects (\moneyTransferManual\xspace cases) transfer assets manually;
only \moneyTransferAutomatic\xspace projects manage part of the money transfers automatically. 
Examples for the latter are 
\href{http://www.amply.tech/}{Amply},
\href{https://www.educategirls.ngo/}{EducateGirls}, 
\href{https://www.giftcoin.org/}{GiftCoin}, and
\href{https://alice.si/}{Alice}, 
which assess the progression of projects and use the fulfillment of goals as a criterion
to trigger payments through smart contracts.

\section{Fund raising}
\label{sec:funding}

In this~\namecref{sec:funding} we study how social good projects get funding.
Overall, \totProjectsFunding out of the \totProjects projects in our collection
have received some funds.
As far as we can tell from the information available online,
only \totProjectsFundingOther of them have been funded through conventional sources,
like private investors or international organizations.
These projects gathered \usdProjectsFundingOther in total,
with a peak of \USD 30M raised by \href{https://www.abra.com/}{Abra},
a financial inclusion project which enables cross-border money remittances.
The other \totProjectsFundingICOnonzero projects have received funds through
Initial Coin Offerings (ICOs).
These are a form of fund raising that has become widespread since 
2017~\cite{Adhami18icos,Zetzsche17ico}.
In an ICO, the project founders
create and put on the market a set of crypto-assets, called \emph{tokens}.
Investors fund the project by buying these tokens,
hoping that they will gain value if the project is successful.
By querying various ICO trackers we have found that
\totProjectsFundingICOany projects (among those in our collection)
have launched, or are expected to launch, an ICO.
Out of the \totProjectsFundingICOclosed projects for which the ICO 
has been closed,
\totProjectsFundingICOnonzero projects have received some money,
for a total of \usdProjectsFundingICO%
\footnote{The information about the money raised by ICOs 
reported by ICO trackers are not always reliable, 
since they are provided by the projects themselves.
However, in most cases this is the only source of information available,
since the addresses used to raise funds are not usually made public.}.

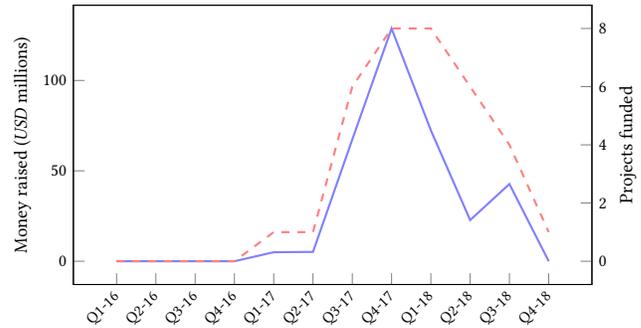
\begin{figure}
  \centering
  \scriptsize
  \scalebox{1}{
    \begin{tikzpicture}
      \begin{axis}[
        height=0.625\linewidth,
        width = 1\linewidth,
        ymin=0,
        ylabel absolute, ylabel style={yshift=0.0cm},
        y label style={at={(0.08,0.5)}},
        ylabel = {Money raised (\USD millions)},
        flexible xticklabels from table={ico-raised-quarter.csv}{Quarter}{col sep=comma},
        xtick=data,
        xtick pos=left,
        nodes near coords align={vertical},
        xticklabel style={font=\tiny}, 
        y tick label style={font=\tiny}, 
        ytick pos=left,
        bar width=5pt,
        xticklabel style = {rotate=45,anchor=east,yshift=-0.1cm},
        enlargelimits = true,
        ymajorgrids = false,
        ]
        \pgfplotstableread[col sep=comma]{ico-raised-quarter.csv}\data
        \addplot [sharp plot,draw=blue!50,solid,thick,mark=none] table[x expr=\coordindex,y=Raised] {\data};
      \end{axis}
      \begin{axis}[
        height=0.625\linewidth,
        width = 1\linewidth,
        ylabel near ticks, 
        yticklabel pos=right,
        ymin=0,
        ylabel = {Projects funded},
        xmajorticks=false,
        nodes near coords align={vertical},
        xticklabel style={font=\tiny}, 
        y tick label style={font=\tiny}, 
        ytick pos=right,
        bar width=5pt,
        xticklabel style = {rotate=45,anchor=east},
        enlargelimits = true,
        ymajorgrids = false,
        ]
        \pgfplotstableread[col sep=comma]{ico-raised-quarter.csv}\data
        \addplot [sharp plot,draw=red!50,dashed,thick,mark=none] table[x expr=\coordindex,y=Projects] {\data};
      \end{axis}
    \end{tikzpicture}
  }
  \vspace{-15pt}
  \caption{Funds raised through ICOs.}
  \label{fig:ico-raised-quarter}
\end{figure}

In~\Cref{fig:ico-raised-quarter} we measure the 
overall funding received by the projects in our collection
through ICOs, showing the temporal evolution by quarters of year 
(solid blue line),
as well as the number of projects funded (dashed red line)%
\footnote{The fall in the number of funded projects in Q3 2018 is probably due to the 
fact that the third quarter is not completed yet at the time of submission
(3rd September, while Q3 will end on 30th September 2018).}.
To associate projects to quarters we consider the average between the 
date on which the ICO was started, and the date on which it was closed.
The peak in Q3 2017 is mainly contributed by 
\USD 20M raised by \href{https://moedaseeds.com/}{Moeda}
(a platform for investing in agriculture projects)
and by \USD 15M raised \href{https://propy.com/}{Propy}
(a project to tokenize land properties),
while the peak in Q3 2018 is mainly due to \USD 42M raised by \href{https://4new.io/}{4NEW}
(a project for producing and distributing green energy).

\begin{figure}[t]
  \scriptsize
  \begin{tikzpicture}
    \begin{axis}[
      legend columns=3, 
      legend style={/tikz/column 2/.style={column sep=5pt,},},
      legend style={at={(0.5,1.3)},anchor=north},
      height=0.625\linewidth,
      width  = 1\linewidth,
      every axis legend/.append style={nodes={right}},
      scaled y ticks = false,
      ylabel absolute, ylabel style={yshift=0.0cm},
      x label style={at={(0.5,0.03)}},
      y label style={at={(0.08,0.5)}},
      x tick label style={/pgf/number format/.cd,fixed,fixed zerofill,precision=0,/tikz/.cd},
      xmin=0,xmax=45,
      ymin=0,ymax=45,
      xminorgrids = true,
      yminorgrids = true,
      xmajorgrids = false,
      ymajorgrids = false,
      xlabel={Money raised (\USD millions)},
      ylabel={Soft cap (\USD millions)},
      nodes near coords,
      ]
      \pgfplotstableread[col sep=comma]{ico.csv}\data
      \addplot[scatter/classes={Y={mark=*,blue},N={mark=triangle*,red}},
      scatter,
      only marks,
      scatter src=explicit symbolic] 
      table[x=Raised,y=SoftCap,meta=Success] {\data};
      \draw [gray,dotted] (rel axis cs:0,0) -- (rel axis cs:1,1);
    \end{axis}
  \end{tikzpicture}
  \vspace{-5pt}
  \caption{Money raised through ICO \emph{vs.} soft cap.}
  \label{fig:ico-raised-vs-softcap}
\end{figure}
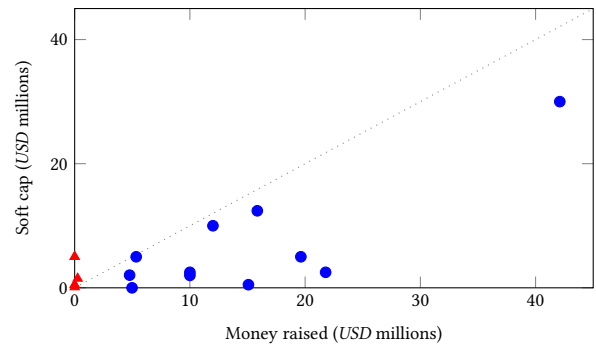

In~\Cref{fig:ico-raised-vs-softcap} we analyse the success of the ICOs,
by comparing the money raised through the ICO and the money expected.
We represent as a point in the diagram 
each of the \totProjectsFundingICOboth projects 
for which we have been able to determine both 
the actual money  raised through the ICO (on the $x$-axis)
and the \emph{soft cap}, 
\ie the money expected by the project founders (on the $y$-axis).
%
The \totProjectsFundingICOsuccess blue circles correspond to the projects for which
the fund raising was successful, \ie the money raised exceeded the soft cap.
Instead, the \totProjectsFundingICOinsuccess red triangles correspond to the unsuccessful ICOs:
in some cases, these project were abandoned just after closing the ICO.

\section{Analysis of blockchain architectures}
\label{sec:architecture}

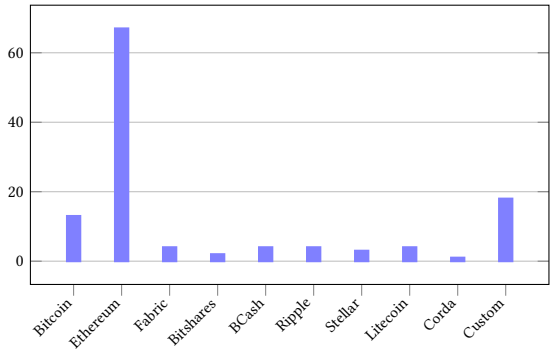
\begin{figure}
  \centering
  \scriptsize
  \scalebox{1}{
    \begin{tikzpicture}
      \begin{axis}[
        height=0.625\linewidth,
        width = 1\linewidth,
        ybar,
        ymin=0,
        flexible xticklabels from table={blockchains.csv}{Blockchain}{col sep=comma},
        xtick=data,
        xtick pos=left,
        nodes near coords align={vertical},
        xticklabel style={font=\tiny}, 
        y tick label style={font=\tiny}, 
        bar width=5pt,
        xticklabel style = {rotate=45,anchor=east},
        enlargelimits = true,
        ymajorgrids = true,
        ]
        \pgfplotstableread[col sep=comma]{blockchains.csv}\data      
        \addplot [fill=blue!50,draw=blue!50,solid,thick,mark=none] table[x expr=\coordindex,y=Amount] {\data};
        \end{axis}
    \end{tikzpicture}
  }
  \caption{Blockchains adopted by the projects.}
  \label{fig:blockchains}
\end{figure}

In this~\namecref{sec:architecture} we analyse the blockchain architectures
of the projects in our collection.
We start by studying which blockchains are used to implement their application logic%
\footnote{We do not consider, in this analysis, the blockchain used for the ICO (in most cases, this is Ethereum).}.
To retrieve this information, we inspect the project websites and whitepapers (when available):
out of the \totProjects projects in our collection, 
we have managed to determine the blockchain used by \totProjectsInfoBlockchain projects;
\totProjectsMoreBlockchain of them use more than one blockchain.
\Cref{fig:blockchains} shows the number of times each blockchain is used by a project.
The most used blockchain is Ethereum (\totProjectsBlockchainEthereum projects);
\totProjectsBlockchainCustom projects develop their own blockchains.

\definecolor{NoBlockchainColor}{RGB}{172,172,172}
\definecolor{PermissionedColor}{RGB}{255,165,0}
\definecolor{PermissionlessColor}{RGB}{173,255,47}

Another relevant question we try to answer is whether the projects are using correctly 
the blockchain technology.
Several processes for deciding, given the project features, which blockchain architecture is needed
are described in literature~\cite{Koens18cbt}.
Among them, in this survey we apply the process in~\cite{Wust18cvcbt}.
This process uses a flow chart (displayed in~\Cref{fig:wust})%
\footnote{Compared to~\cite{Wust18cvcbt}, our flow chart is a bit simplified.
First, we omit the first decision (``Do you need to store a state?''), since the answer is 
always yes for the projects in our collection.
We also drop the decision ``Is public verifiability required?'',
which allows to distinguish between public and private (permissioned) blockchains,
since the information about the blockchain used by our projects 
do not allow to distinguish between these two cases.}
to determine, given a set of basic architectural choices, 
which kind of blockchain (if any) is appropriate for the given use case.
The flow chart has 3 possible outcomes: no blockchain,
\emph{permissioned} or \emph{permissionless} blockchain.
If there exists only one entity authorized to update the blockchain (\ie, a single writer), 
or multiple writers which are either all trusted
or can be safely replaced by a trusted third party which is always online,
then there is no need for a blockchain.
In the other case using a blockchain is appropriate: 
this should be permissioned if all writers are trusted, 
otherwise permissionless. 
Indeed, in a permissioned blockchain (\eg, Hyperledger Fabric) there exists an entity that grants permissions, 
for instance it decides which is the set of users authorized to append new transactions to the blockchain. 
Instead, in a permissionless blockchain (\eg, Ethereum and Bitcoin) there is no such authority.

\tikzstyle{decision} = [diamond, draw, fill=blue!20,text width=6em, text badly centered, node distance=4cm, inner sep=0pt]
\tikzstyle{block} = [rectangle, draw, fill=blue!20,text width=6em, text centered, rounded corners, minimum height=4em]
\tikzstyle{line} = [draw, -latex']

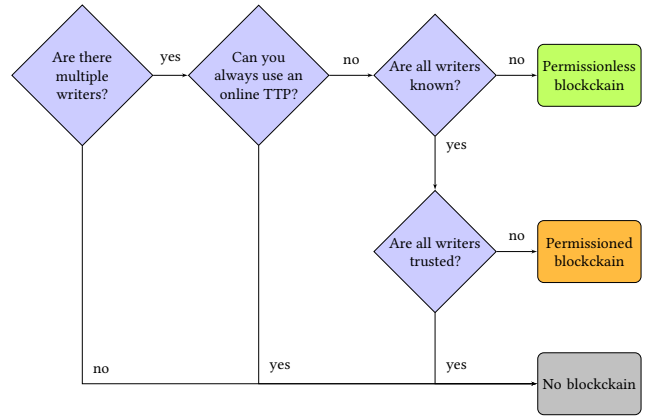
\begin{figure}    
  \resizebox{\linewidth}{!}{
    \begin{tikzpicture}[node distance = 8cm]
      \node [decision] (multipleWriters) {Are there multiple writers?};
      \node [decision, right of=multipleWriters] (onlineTTP) {Can you always use an online TTP?};
      \node [decision, right of=onlineTTP] (writersKnown) {Are all writers known?};
      \node [decision, below of=writersKnown] (writersTrusted) {Are all writers trusted?};
      \node [block, right of=writersKnown, node distance=3.5cm, fill=PermissionlessColor!75] (permissionless) {Permissionless blockckain};
      \node [block, right of=writersTrusted, node distance=3.5cm, fill=PermissionedColor!75] (permissioned) {Permissioned blockckain};
      \node [block, below of=permissioned, node distance=3cm, fill=NoBlockchainColor!75] (noblockchain) {No blockckain};
      \path [line] (multipleWriters) -- node[label=above:yes] {} (onlineTTP);
      \path [line] (multipleWriters) |- node[label=above right:no] {} (noblockchain);
      \path [line] (onlineTTP) -- node[label=above:no] {} (writersKnown);
      \path [line] (onlineTTP) |- node[label=above right:yes] {} (noblockchain);
      \path [line] (writersKnown) -- node[label=above right:yes] {} (writersTrusted);
      \path [line] (writersKnown) -- node[label=above:no] {} (permissionless);
      \path [line] (writersTrusted) |- node[label=above right:yes] {} (noblockchain);
      \path [line] (writersTrusted) -- node[label=above:no] {} (permissioned);
    \end{tikzpicture}
  }
  \caption{A flow chart to determine which blockchain technology is appropriate for a given use case (inspired to~\cite{Wust18cvcbt}).}
  \label{fig:wust}
\end{figure}

\tabulinesep=1.5mm

The following table displays the results of our analysis.
For \totProjectsActualDiffersPredicted projects 
the actual blockchain does not coincide with the predicted one:
for instance \totProjectsArchitecturePermissionlessPermissioned 
projects use a permissionless blockchain (almost always, Ethereum), 
whereas they should have used a permissioned one.
This choice may negatively impact the effectiveness of the project,
\eg requiring higher fees for appending transactions or running smart contracts.
\vspace{-5pt}
\begin{center}
  \resizebox{\columnwidth}{!}{
    \begin{tabu}{l|c|c|c|c|}
      \multicolumn{2}{c}{} & \multicolumn{3}{c}{Predicted}
      \\
      \cline{3-5}
      \multicolumn{1}{c}{}& & \textbf{Permissioned} & \textbf{Permissionless} & \textbf{No} 
      \\
      \cline{2-5}
      \multirow{2}{*}{\rotatebox{90}{Actual}} 
      & \textbf{Permissioned} & 4 & 2 & 0 
      \\
      \cline{2-5} 
      & \textbf{Permissionless} & 5 & 13 & 1 
      \\
      \cline{2-5}
    \end{tabu}
  }
\end{center}

\section{Status of projects}
\label{sec:project-status}

In this~\namecref{sec:project-status} we analyse the evolution of projects,
to find how many of them remain in the status of proposal, 
how many become operational, and how many are eventually abandoned.
To this purpose we associate each project in our collection with one of the following tags:
\begin{description}
\item[Proposal] the project has been advertised online, 
  and it may have issued an ICO to gather funding (the ICO may be either upcoming or closed).
  However, the project has not started the development lifecycle.
\item[Prototype] the project is not operational yet, 
  but its development lifecycle has started:
  some prototype, demo, or open-source repository is available to the public.
\item[Live] the project is operational and available to final users.
\item[Abandoned] the project website is down, or it displays obsolete content,
  or there is some external source which declares the project abandoned.
\end{description}

\begin{figure}
  \begin{tikzpicture}
    \begin{axis}[
      height=0.625\linewidth,
      width=\linewidth,
      x tick label style={/pgf/number format/1000 sep=},
      flexible xticklabels from table={project-status.csv}{Year}{col sep=comma},
      xtick=data,
      xtick pos=left,
      enlargelimits=0.05,
      legend style={
        at={(0.175,0.95)}, 
        anchor=north,
        legend columns=1,
        font=\tiny,
      },
      xmin=2010.7,
      xmax=2018.3,
      ybar,
      bar width=3pt,
      xticklabel style={font=\tiny}, 
      yticklabel style={font=\tiny}, 
      ]
      \pgfplotstableread[col sep=comma]{project-status.csv}\mytable
      \addplot[red,fill=red!50] table [y index=1,x=Year] {\mytable};
      \addplot[blue,fill=blue!50] table [y index=2,x=Year] {\mytable};
      \addplot[green,fill=green!50] table [y index=3,x=Year] {\mytable};
      \addplot table [y index=4,x=Year] {\mytable};
      \legend{Proposal,Prototype,Live,Abandoned}
    \end{axis}
  \end{tikzpicture}
  \vspace{-5pt}
  \caption{Status of projects by year of proposal.}
  \label{fig:project-status}
\end{figure}
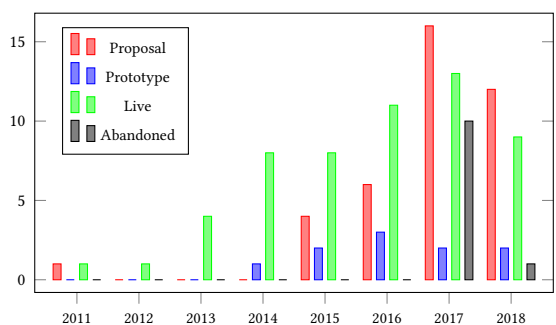

To keep track of the status of projects we use the information available online,
while we do not consider the ``development roadmap'' often included in the project whitepapers,
since it may not correspond to reality.
\Cref{fig:project-status} shows the status of projects,
grouping them by the year when they were proposed.
There is a peak of proposals in 2017 
(probably related to the overall growth of Ethereum-based projects),
with a corresponding peak of abandoned projects%
\footnote{As before, the fall of 2018 may be due to missing data at time of writing.}.

We also count how many projects are developed in the open-source environment.
Indeed, open-source is a common trait of blockchain technologies:
the survey~\cite{Deloitte17github} counts \mbox{$\sim$86K} open-source projects based on blockchain (not only for social good) 
up to October 2017, with an average of \mbox{$\sim$8K} new projects each year.
Similarly to~\cite{Deloitte17github}, we conduct our search on GitHub;
we find that only \totProjectsGithub out of the \totProjects projects in our collection
have published some code on GitHub.
The analysis in~\cite{Deloitte17github} found that only \mbox{8\%} of projects 
in their dataset are \emph{alive}
--- meaning that their GitHub repositories have been updated at least once in the last six months.
A similar analysis on the \totProjectsGithub open-source projects in our collection
yields more positive results: 
\totProjectsGithubAlive of them are still alive (\percentProjectsGithubAlive of the total).

\Cref{tab:comparison} 
relates the availability of documentation and open-source repository (and its liveness) 
to that of funding.
Since values of funded and not funded projects are quite close, 
we cannot infer that funding is related to 
availability of source code (or its liveness) and documentation.

\begin{table}[t]
  \centering
  \resizebox{\linewidth}{!}{%
    \begin{tabular}{|c|c|c|c|}
    	\hline
    	     \textbf{Features}      & \textbf{Funded} & \textbf{Not funded} &                          \textbf{Total}                           \\ \hline
    	        Whitepaper          &       27        &         35          &                      \totProjectsWhitepaper                       \\
    	      GitHub (alive)        &     15 (9)      &       23 (14)       &           \totProjectsGithub (\totProjectsGithubAlive)            \\
    	Whitepaper + GitHub (alive) &     15 (9)      &        13 (9)       & \totProjectsGithubWhitepaper (\totProjectsFundedGithubWhitepaper) \\ \hline
    \end{tabular}%
  } %
  \caption{Projects documentation and code \emph{vs.} funding.}
  \label{tab:comparison}
  \vspace{-10pt}
\end{table}

\section{Conclusions}
\label{sec:conclusions}

Our paper gives a snapshot on factual data about social good projects.
Our dataset represents a significant sample 
of how blockchain technologies are applied for the social good.
Of course our results are not conclusive, since many more years
will be needed to have a clear understanding of how (and if) blockchain technologies will enable social innovation.
However, ours is first step of a long-term effort towards
a systematic analysis of this phenomenon, which we will continue to study
as new data become available.

\subsubsection*{Acknowledgments.} 

Work partially supported by R.A.S.\ projects 
``Sardcoin'' and ``Smart collaborative engineering''.

\bibliographystyle{ACM-Reference-Format}
\bibliography{main}

\end{document}